\documentclass[10pt,twocolumn,american,notitlepage]{revtex4-2}
\usepackage{mathptmx}
\usepackage{helvet}
\usepackage{courier}

\usepackage[LGR,T1]{fontenc}
\usepackage[latin9]{inputenc}
\usepackage{color}
\usepackage{babel}
\usepackage{amsmath}
\usepackage{graphicx}
\usepackage{esint}
\usepackage[pdfusetitle,
 bookmarks=true,bookmarksnumbered=false,bookmarksopen=false,
 breaklinks=true,pdfborder={0 0 1},backref=false,colorlinks=true]
 {hyperref}



\begin{document}
\title{Magnetic Pumping: from Plasma Heating to Particle Acceleration }
\author{Mikhail Malkov }
\affiliation{University of California, San Diego, CA, USA; Eureka Sci., Oakland,
CA, USA}
\author{Immanuel Jebaraj}
\affiliation{Department of Physics and Astronomy, University of Turku, FI-20014
Turun yliopisto, Finland}
\begin{abstract}

One of the earliest mechanisms proposed for plasma heating was magnetic pumping (MP). However, its significance for astrophysical phenomena, including particle acceleration, has yet to be appreciated. MP-energized particles tap energy from magnetic-field oscillations. A particle's momentum component perpendicular to the local B-field increases during field growth by virtue of the adiabatic invariant $p_{\perp}^{2}/B \approx \mathrm{const}$. The gained $p_{\perp}$ is then partially scattered elastically into the parallel momentum, $p_{\parallel}$, with $p^{2}=p_{\parallel}^{2}+p_{\perp}^{2}\approx \mathrm{const}$, thereby retaining some fraction of the gained energy before the field decreases to its minimum. This scattering breaks the reversibility of energy exchange between particles and oscillating magnetic fields, thereby increasing the particle energy after each MP cycle. Field oscillations are often assumed to be sinusoidal, and the resulting MP is treated perturbatively. These simplifications restrict astrophysical applications, leaving objects with strong magnetic perturbations outside the scope of adequate treatment. We develop a nonperturbative approach to MP that is suitable for a broad spectrum of magnetic turbulence. The treatment comprises two steps. The first step is common: converting a kinetic equation into an infinite hierarchy of moments of the particle distribution function. The second step is new in MP treatments: we find an exact closure at an arbitrary level of the moment system. The heating is treated exactly at the second-moment closure, using a first-order ordinary differential (Riccati) equation. Particle acceleration generally requires a higher-level closure to determine the power-law index and the maximum energy of accelerated particles. However, we propose a method for extracting these crucial acceleration data from the second moment for a broadband random field.

\end{abstract}
\maketitle

\section{Introduction\protect\label{sec:Introduction}}

Magnetic pumping (MP) is a plasma heating mechanism in which time-dependent magnetic compressions and rarefactions drive reversible changes in particle momenta through conservation of adiabatic invariants, while pitch-angle scattering breaks this reversibility and produces net energization. In its simplest form, if the magnetic field strength $B\left(t\right)$ varies slowly compared to the particle gyroperiod, the first adiabatic invariant $p_{\perp}^{2}/B$ is approximately conserved, so the perpendicular momentum $p_{\perp}\left(t\right)$ follows $B\left(t\right)$ along particle trajectories (see, e.g., early treatments in \citealp{BergerNewcomb1958} and later discussions in \citealp{Borovsky1986,Lichko2017,Ley2023}). These reversible $p_{\perp}$ variations generate pitch-angle anisotropy in the distribution, which relaxes in the presence of scattering. Because scattering transfers part of the perpendicular excess into the parallel degree of freedom during the compression phase, the energy gained at large $B\left(t\right)$ is not fully returned during the subsequent rarefaction, leading to secular growth of $p^{2}=p_{\parallel}^{2}+p_{\perp}^{2}$. 

In laboratory plasmas the required scattering may be collisional e.g., \citep{LaroussiRoth1989}, whereas in most space and astrophysical environments it is more naturally provided by wave--particle interactions, self-generated microinstabilities, and magnetic turbulence (e.g., \citealp{Borovsky1986,Lichko2017,Ley2023}). The anisotropy produced by MP can also excite fluctuations that contribute to the scattering, potentially forming a feedback loop (e.g., \citealp{Lyutikov2007,Ley2023}).

In this paper, we develop a nonperturbative analysis of MP heating and extends it toward the acceleration of nonthermal particles. The main technical advance is an exact closure of the (infinite) moment hierarchy of the kinetic equation at an arbitrary moment order, rather than truncating the hierarchy at low order or relying on small-amplitude perturbative expansions. This exact closure is designed to remain valid when magnetic/compressive fluctuations are strong and spectrally broad. Conditions that are often expected, and in many cases observed, in astrophysical plasmas. MP has been invoked in a wide range of such settings, including pulsar nebulae \citep{Melrose1969}, relativistic jets \citep{MalkovICRC_Jet2023}, the intracluster medium \citep{Lyutikov2007,Ley2023}, stellar winds and the solar wind \citep{Lichko2017}, interplanetary shocks \citep{Kennel1986,Jebaraj2024ApJL}, planetary magnetospheres and radiation belts \citep{Borovsky1981,Borovsky2017,GolanGedalin2025}, and reconnecting magnetic-field configurations \citep{Drake_2013}. In these environments, large-amplitude fluctuations and broad spectra can limit the applicability of strictly perturbative treatments. While numerical simulations are indispensable, they are necessarily restricted in parameter range and can be sensitive to numerical heating (e.g., \citealp{Ley2023}). Given continuing progress in in situ and remote observations, it is timely to develop analytic results that are both less restrictive and suitable for code verification; complementary single-point diagnostics of energization mechanisms have also been developed in recent years (e.g., field--particle correlation analyses applied to MP; \citealp{MontagHowes2022}).

Historically, MP was devised as a heating concept for laboratory magnetic-fusion plasmas in the 1950s, including a classified study by Spitzer and Witten (Princeton AEC report; see the historical discussion and related collisional-pumping theory in \citep{LaroussiRoth1989}, and the broader review in \citealp{Borovsky1986}). Early open-literature formulations of pumping-based heating include \citet{BergerNewcomb1958}, and subsequent variants and optimizations were proposed in the fusion context (e.g., \citealp{KoechlinSamain1971}). In the magnetic-fusion setting, rapid advances later shifted emphasis toward other heating techniques, and MP received comparatively less attention. In contrast, renewed interest in MP has been driven by astrophysical and space-plasma applications, where compressions, shocks, and turbulence are ubiquitous and where scattering is often provided naturally by electromagnetic fluctuations rather than by Coulomb collisions (e.g., \citealp{Borovsky1986,Lichko2017,Ley2023}). The present work is motivated by these regimes and aims to place MP on a footing that does not rely on small fluctuation amplitudes or narrowband driving.

A central motivation for a nonperturbative theory is that the commonly imposed condition $\Delta B/B\ll1$ is frequently violated in astrophysical and geophysical plasmas. In many systems the opposite ordering, $\Delta B\gg B_{0}$ (where $B_{0}$ is a representative background field), can apply. For example, magnetic-field strength surges at collisionless shock ramps and can remain strongly perturbed downstream; upstream, shock-accelerated particles can excite waves via cyclotron resonance and amplify them through current-driven \citep{Bell04} and pressure-driven instabilities (see, e.g., \citealp{DruryFal86,MDS10PPCF,BykBrandMalk13}). In heliospheric contexts, large-amplitude, multi-scale compressions and turbulence near shocks and in their foreshocks are routinely implicated in particle energization, and MP has been explicitly advanced as a seed/heating mechanism in the vicinity of Earth's bow shock \citep{LichkoEgedal2020}. These considerations motivate treating MP for arbitrary $\Delta B/B$, particularly when addressing particle heating and acceleration in strongly compressive environments.

The small-amplitude restriction may be acceptable in comparatively quiescent solar-wind conditions (e.g., \citealp{Lichko2017}), but it becomes problematic in more structured flows such as corotating interaction regions and turbulent stream interfaces, especially where reconnection occurs \citep{Drake_2013}, and it is even less suitable in shock environments where heating and acceleration are central problems. Moreover, in strong but short-scale magnetic perturbations, energetic particles may still experience only small net deflections because of their large Larmor radii, despite large field variations. This can place transport and energization in a regime that differs from the standard quasilinear picture based on small perturbations and resonant diffusion. In particular, the quasilinear approximation has been argued to be inconsistent with flat spectra often inferred near some interplanetary shocks \citep{Malkov2024a}. Within this context, MP has been proposed as a mechanism capable of producing localized pockets of energetic particles observed upstream of certain interplanetary shocks, including cases where particle fluxes appear nearly energy independent over limited energy ranges \citep{Perri2023}. Independent evidence that compressions and associated anisotropy-driven scattering can control irreversible electron heating at collisionless shocks has also been found in kinetic (PIC) studies, where shock-amplified fields drive temperature anisotropy and whistler turbulence (\citealp{Guo2017ApJ}; see also recent applications in trans-relativistic shocks \citealp{SironiTran2024}).

A second common simplification we relax is the use of a relaxation-time ($\tau$) representation for pitch-angle scattering, which replaces the scattering operator by $\left(\left\langle f\right\rangle -f\right)/\tau$, where $\left\langle \cdot\right\rangle $ denotes pitch-angle averaging and $\tau$ is an assumed isotropization time. This representation is generally not appropriate for small-angle deflections typical of interactions with random-phase wave ensembles, which are central to the broadband-driving regime considered here. We instead employ the Lorentz (pitch-angle diffusion) operator (see, e.g., standard cosmic-ray/space-plasma treatments and numerical tests in \citealp{Tautz2013}),
\[
\partial_{\mu}\left(1-\mu^{2}\right)\partial_{\mu}f
\]
where $\mu$ is the pitch-angle cosine. The principal methodological contribution of this work is then to close, exactly and at arbitrary order, the resulting hierarchy of moment equations for the particle distribution function. For plasma heating, closure at the second moment is often sufficient. For nonthermal particle acceleration, higher moments are generally needed to represent the emergence and evolution of extended power-law tails. Nevertheless, we show that for broadband, high-amplitude pumping, the acceleration dynamics can be captured to useful accuracy with a limited number of moments, providing a practical and systematically improvable analytic framework for interpreting observations and benchmarking simulations.

The remainder of the paper is organized as follows. In Sec.~\ref{sec:General-Magnetic-Pumping} we formulate magnetic pumping starting from a drift-kinetic description with pitch-angle diffusion and derive the corresponding moment hierarchy together with its exact finite closure. In Sec.~\ref{subsec:Plasma-Heating} we apply the lowest-order closure to obtain a compact, nonperturbative description of plasma heating and evaluate standard approximations against an exact Floquet treatment for periodic and broadband drivers. In Sec.~\ref{subsec:Particle-Acceleration} we extend the moment-closure framework toward nonthermal particle acceleration and characterize how broadband driving controls energization efficiency. Finally, in Sec.~\ref{sec:Discussion} we summarize the main results, discuss limitations and applicability, and outline implications for observations and numerical tests.

\section{Magnetic Pumping Model\protect\label{sec:General-Magnetic-Pumping}}

Existing treatments of MP, some of which were referenced in the Introduction,
often differ in the specific approximations adopted for the underlying particle dynamics.
To clarify the assumptions and notation used in this work, we start from the standard
drift-kinetic equation (DKE) for the particle distribution function
$f\left(t,\boldsymbol{r},p_{\parallel},p_{\perp}\right)$, in which only the fast gyromotion
is averaged out.

\subsection{Drift-Kinetic equation (DKE)\protect\label{subsec:Drift-Kinetic-equation-(DKE)}}

In the next section, using the DKE introduced below, we will distinguish between three physical situations and unify them in a common mathematical form. At the basic level, the DKE may still be viewed as a Liouville equation expressing conservation of phase volume along particle trajectories, $df/dt=0$. Here, however, the time derivative is taken along the guiding-center (GC) trajectory in the reduced, 5-D phase space $\left(\boldsymbol{r},p_{\parallel},p_{\perp}\right)$, rather than along the full 6-D particle trajectory \citep{Northrop1963}. We then add to the right-hand side a collision term, $\text{St}f$, that accounts for pitch-angle scattering. As discussed in the Introduction, the collision term may represent Coulomb collisions with other charged and neutral particles, or scattering by waves and fluctuations associated with magnetic perturbations. These effects are not included in the explicit acceleration terms $\dot{p}_{\parallel,\perp}$ on the left-hand side of eq.~(\ref{eq:DKE-1}) below, but are instead collected into the operator $\text{St}f$. Various equivalent forms of the DKE appear in the literature (e.g., \citep{Sivukhin1965} for a review); at the level needed here they can be written as

\begin{equation}
\frac{\partial f}{\partial t}+\dot{\boldsymbol{r}}\cdot\nabla f+\dot{p}_{\parallel}\frac{\partial f}{\partial p_{\parallel}}+\dot{p}_{\perp}\frac{\partial f}{\partial p_{\perp}}=\text{St}f\label{eq:DKE-1}
\end{equation}

Mathematically involved expressions for the functions $\dot{\boldsymbol{r}}\left(\boldsymbol{r},p_{\parallel},p_{\perp}\right)$, $\dot{p}_{\parallel}\left(\boldsymbol{r},p_{\parallel},p_{\perp}\right)$, and $\dot{p}_{\perp}\left(\boldsymbol{r},p_{\parallel},p_{\perp}\right)$ can be found in the above reference, and in a somewhat more compact form in \citep{Morozov1966}. We will not require their explicit forms, except for the following relation, which can be directly obtained from \citep{Sivukhin1965}:

\begin{equation}
\nabla\cdot\dot{\boldsymbol{r}}+\frac{\partial\dot{p}_{\parallel}}{\partial p_{\parallel}}+\frac{\partial\dot{p}_{\perp}^{2}}{\partial p_{\perp}^{2}}=0.\label{eq:DKE-ID}
\end{equation}

This identity allows eq.~(\ref{eq:DKE-1}) to be written in conservative form,

\begin{equation}
\frac{\partial f}{\partial t}+\nabla\cdot\left(f\dot{\boldsymbol{r}}\right)+\frac{\partial}{\partial p_{\parallel}}\left(\dot{p}_{\parallel}f\right)+\frac{1}{p_{\perp}}\frac{\partial}{\partial p_{\perp}}\left(p_{\perp}\dot{p}_{\perp}f\right)=\text{St}f,\label{eq:DKE-ConsForm}
\end{equation}

assuming, of course, that $\text{St}f$ also conserves the number of particles (and energy). In many MP applications the explicit spatial dependence of $f\left(\boldsymbol{r}\right)$ is not essential, so we may average over $\boldsymbol{r}$, assuming that there is no significant flux $f\dot{\boldsymbol{r}}$ through the boundaries of the domain considered. This flux vanishes identically under periodic boundary conditions, which are frequently employed in simulations. Under such averaging, the second term on the left-hand side vanishes. However, a closure problem arises for the terms $\overline{\dot{p}_{\parallel,\perp}f}$ (the overbar denotes spatial averaging). For a statistically homogeneous turbulent plasma, we may treat $f$ as coordinate independent, or assume that any residual coordinate dependence is negligible compared to the momentum-space evolution driven by $\dot{p}_{\parallel,\perp}$. In that case we approximate $\overline{\dot{p}_{\parallel,\perp}f}\approx\overline{\dot{p}}_{\parallel,\perp}\,\overline{f}$. Any weak coordinate dependence of these quantities can be retained in the resulting equation as a parameter, but it will be ignored in the remainder of this paper, and the overbars will be dropped. The averaged equation then reduces to

\begin{equation}
\frac{\partial f}{\partial t}+\frac{\partial}{\partial p_{\parallel}}\left(\dot{p}_{\parallel}f\right)+\frac{1}{p_{\perp}}\frac{\partial}{\partial p_{\perp}}\left(p_{\perp}\dot{p}_{\perp}f\right)=\text{St}f.\label{eq:DKE-ConsAver}
\end{equation}

We now review the three major models for $\dot{p}_{\parallel,\perp}$.

\subsection{Plasma Heating and Particle Acceleration: Unified Model for $\dot{p}_{\parallel}$ and $\dot{p}_{\perp}$ under MP\protect\label{subsec:Heating-and-Accel}}

Below, we consider three regimes of particle energization by fluctuating magnetic fields. We start with their common characteristics and turn to a specific treatment of eq.~(\ref{eq:DKE-ConsAver}) in the next subsection. Since, in a slowly varying field $B\left(t\right)$, particles conserve the adiabatic invariant $J_{\perp}=p_{\perp}^{2}/B$ with exponential accuracy, we adopt for $\dot{p}_{\perp}$ in eq.~(\ref{eq:DKE-ConsAver}) the standard relation
\begin{equation}
\dot{p}_{\perp}=p_{\perp}\frac{\dot{B}}{2B}.\label{eq:p-perp}
\end{equation}
Let us further assume that the second adiabatic invariant, $J_{\parallel}=\oint p_{\parallel}dl$, is also conserved. Here $dl$ is the length element of the GC orbit along the field. Conservation of $J_{\parallel}$ requires that the particle motion along the field be either finite (as in a magnetic trap) or periodic (as in a toroidal magnetic field or a periodic wave field). Note that, in general, the particle orbit need not be periodic even when the field is. To complete the list of adiabatic invariants, if the GC also drifts across the field in a nearly periodic fashion, the third invariant corresponds to the magnetic flux through the GC orbit projected onto the plane perpendicular to the field. We will not invoke the third invariant here.

While $\dot{p}_{\perp}$ is fixed by eq.~(\ref{eq:p-perp}), $\dot{p}_{\parallel}$ must be obtained from conservation of $J_{\parallel}$ in a manner that depends on the system configuration. In general guiding-center theory, the evolution of $p_{\parallel}$ can be written as $\dot{p}_{\parallel}=eE_{\parallel}-\mu_{B}\nabla_{\parallel}B$ \citep{Morozov1966}, where $\mu_{B}=v_{\perp}p_{\perp}/2B$ is the particle magnetic moment. If the conductivity along the field can be regarded as effectively infinite, the contribution of the parallel electric field, $E_{\parallel}$, to $\dot{p}_{\parallel}$ can be neglected. If, in addition, the spatial variation of the magnetic field does not contribute significantly to $\overline{\dot{p}_{\parallel}f}$ in the averaged eq.~(\ref{eq:DKE-ConsForm}), we arrive at case (1): $\dot{p}_{\parallel}=0$. In this case, $p_{\parallel}$ changes only through scattering, subject to the constraint $p_{\parallel}^{2}+p_{\perp}^{2}\equiv p^{2}=\mathrm{const}$.

The remaining two cases (2,3) are as follows (see, e.g., \citep{Borovsky1986,Ley2023,Lichko2017} for more discussion). In case (2), a \emph{conserved plasma volume} is enclosed within a magnetic flux tube with variable length and cross section $S$, so that $V=\oint S\,dl=\mathrm{const}$. Since the magnetic flux through the tube, $BS$, is constant along the tube, we can rewrite the volume as
\[
V=BS\oint\frac{dl}{B}=\mathrm{const}.
\]
Hence, if both $V$ and $J_{\parallel}=\oint p_{\parallel}dl$ are conserved, then $p_{\parallel}\propto B^{-1}$. In case (3), the total number of particles in the tube is conserved rather than its volume. In that case,
\[
N=SB\oint n\,\frac{dl}{B}=\mathrm{const},
\]
where $n$ is the plasma density, so that $p_{\parallel}\propto n/B$, leading to $\dot{p}_{\parallel}\approx p_{\parallel}\left(\dot{n}/n-\dot{B}/B\right)$. Defining, for case (3), the function $\sigma\left(t\right)\equiv2\left(\dot{B}/B-\dot{n}/n\right)/\left(\dot{B}/B\right)$, we can combine all three cases into the unified expression
\begin{equation}
\dot{p}_{\parallel}=-\sigma p_{\parallel}\dot{B}/2B\label{eq:p-paral}
\end{equation}
where $\sigma=0$ for case (1) ($\dot{p}_{\parallel}=0$) and $\sigma=2$ for case (2) provided the motion along the field is finite (periodic). For case (3), $N=\mathrm{const}$, it is convenient to use the equivalent form $\sigma\left(t\right)\equiv2-2\,\partial\ln n/\partial\ln B$. We collect these options as
\begin{equation}
\sigma=\begin{cases}
0, & E_{\parallel}\approx\nabla_{\parallel}B\approx0\\
2, & V=\mathrm{const}\\
2-2\partial\ln n/\partial\ln B, & N=\mathrm{const}
\end{cases}\label{eq:3Sigmas}
\end{equation}
They can be unified by noting that the first and second cases are special cases of the third for $n\propto B$ and $n=\mathrm{const}$, respectively. Using the above expressions for $\dot{p}_{\parallel}$ and $\dot{p}_{\perp}$, in the next subsection we will transform eq.~(\ref{eq:DKE-ConsAver}) into a hierarchy of moment equations for the Lorentz gas scattering operator $\mathrm{St}f$.

\subsection{Moment Description of Magnetic Pumping\protect\label{subsec:Moment-Description-of}}

Now that we have rules for computing $\dot{p}_{\perp}$ and $\dot{p}_{\parallel}$, given by eqs.~(\ref{eq:p-perp}) and (\ref{eq:p-paral}), we turn to specifying the right-hand side of eq.~(\ref{eq:DKE-ConsAver}). For astrophysical applications of MP, a suitable yet simple scattering model for the $\mathrm{St}f$ term in eq.~(\ref{eq:DKE-ConsAver}) is the Lorentz gas approximation, $\mathrm{St}f=\nu\,\partial_{\mu}\left(1-\mu^{2}\right)\partial_{\mu}f$. Since the scattering frequency $\nu$ is assumed to be momentum independent, we use it as a time unit by replacing $\nu t\to t$. We also introduce a new dependent variable $F=p^{2}f\equiv\left(p_{\parallel}^{2}+p_{\perp}^{2}\right)f$. Equation~(\ref{eq:DKE-ConsAver}) then rewrites as follows:

\begin{eqnarray}
\frac{\partial F}{\partial t}+\frac{\dot{B}}{2B}\left\{ \frac{\partial}{\partial p}p\left[1-\left(1+\sigma\right)\mu^{2}\right]F\right. & -\nonumber \\
\left.\left(1+\sigma\right)\frac{\partial}{\partial\mu}\mu\left(1-\mu^{2}\right)F\right\}  & = & \frac{\partial}{\partial\mu}\left(1-\mu^{2}\right)\frac{\partial F}{\partial\mu}\label{eq:F}
\end{eqnarray}

Here $F\left(t,p,\mu\right)$ is normalized to $dp\,d\mu$ and depends on the variables $p=\sqrt{p_{\parallel}^{2}+p_{\perp}^{2}}$ and $\mu=p_{\parallel}/\sqrt{p_{\parallel}^{2}+p_{\perp}^{2}}$ in place of $p_{\parallel}$ and $p_{\perp}$. Next, we convert the above PDE to an infinite system of ODEs for the moments,
\begin{equation}
M_{ij}=\frac{1}{2}\int_{-1}^{1}\mu^{i}d\mu\int_{0}^{\infty}p^{j}Fdp.\label{eq:MomDef}
\end{equation}
The following matrix equation for the moments results:

\begin{eqnarray}
\frac{\partial M_{ij}}{\partial t}+\frac{\dot{B}}{2B}\left\{ \left(j-i\right)\left(1+\sigma\right)M_{i+2,j}+\left[i\left(1+\sigma\right)-j\right]M_{ij}\right\} \nonumber \\
=i\left(i-1\right)M_{i-2,j}-i\left(i+1\right)M_{ij}.\qquad\label{eq:MomEq}
\end{eqnarray}

There are two observations to make about these equations that, to our knowledge, have not been emphasized in the MP literature. First, only matrix elements from the same column, labeled by the index $j$, appear in each triad of equations coupling $M_{ij}$ to $M_{i\pm2,j}$. This property follows from momentum-independent pitch-angle scattering, which is also assumed in many of the preceding MP models discussed in the Introduction. The second, and most important, observation is as follows. We set $M_{-2,j}=M_{-1,j}=0$ as extraneous moments associated with the angular decomposition in eq.~(\ref{eq:MomDef}) (cf., e.g., a Legendre-polynomial expansion). We then fix any column number $j$ and resolve the subsystem for $M_{ij}$ starting from either $i=0$ (if $j$ is even) or $i=1$ (if $j$ is odd). As the row index $i$ increases and reaches $i=j$, the resulting set of equations closes by itself: the coefficient of the next unknown element $M_{i+2,j}$ on the left-hand side of eq.~(\ref{eq:MomEq}) vanishes at this step, while $M_{i-2,j}$ on the right-hand side was already included at the preceding level $i=j-2$. The total number of equations is $j/2+1$ (even $j$) or $\left(j+1\right)/2$ (odd $j$), which equals the number of unknown moments, $M_{1,j},M_{3,j},\dots,M_{j,j}$ or $M_{0,j},M_{2,j},\dots,M_{j,j}$, respectively.

However, the cases of odd column numbers $j$ in the matrix $M$ are physically distinct from the cases of even $j$. Since the MP driver is symmetric in $\mu$, it is natural to expect that odd moments in $\mu$ are not directly driven and therefore decay under pitch-angle scattering. We demonstrate and discuss this property in Appendix~\ref{sec:Odd-Moments}. A special significance is attached to the odd moment $M_{11}$, which decays as $M_{11}\propto e^{-2t}$ irrespective of the driver's amplitude and spectral characteristics. For a nonrelativistic part of the particle population (thermal particles, in practical terms), $M_{11}$ corresponds to the field-aligned current carried by a given species. Its damping reduces MP energy dissipation through the work done per unit time by the parallel electric field on the plasma, $j_{\parallel}E_{\parallel}$.

The strong damping of odd moments also helps formalize the relation between the original drift-kinetic equation and the infinite system of ODEs for the moments. Because odd moments are damped, we may restrict the class of initial conditions to $M_{2i+1,2k+1}\left(0\right)\equiv0$ for all $i,k=0,1,2,\dots$. According to eq.~(\ref{eq:MomEq}), these moments then remain zero for $t>0$. In Appendix~\ref{sec:Odd-Moments} we also consider the evolution of moments in this group for $i=k=0$ and for $i=0,1$ with $k=1$ when they are not zero at $t=0$, and demonstrate that they decay either unconditionally (the case of $M_{11}$) or rapidly (the cases of $M_{13}$ and $M_{33}$), even for a strong driver. We therefore focus in this paper on the even columns of the matrix $M$. Note, however, that resolving the closed subsystems $M_{0,2n},M_{2,2n},\dots,M_{2n,2n}$ (where $n=0,1,\dots$) leaves the remaining moments $M_{2k,2n}$ with $k>n$ in each such column unresolved.

In effect, we neglect the unessential odd moments in $\mu$ by focusing on the exactly closable part of the system in eq.~(\ref{eq:MomEq}), which is also the most relevant physically. Regarding the equivalence of the DKE and its moment description, it is sufficient to choose initial conditions with $F\left(\mu,t=0\right)$ symmetric in $\mu$. By the symmetry of the parent equation~(\ref{eq:F}), this symmetry of $F$ in $\mu$ is then preserved for $t>0$.

Even though the conversion is \emph{exact} for the most important subclass of moments $M_{ij}$ with even $i$ and $j$, and can be extended to arbitrarily large $j$, there is no full equivalence between eq.~(\ref{eq:F}) and its moment representation in eq.~(\ref{eq:MomEq}). Indeed, without a \emph{truncation}, and in addition to the limitations associated with odd moments discussed above, our exact resolution algorithm for the matrix $M_{ij}$ generates only its upper triangle with $i\le j$, even if odd moments are neglected altogether. Nevertheless, this information is sufficient for heating and acceleration purposes, as these processes are characterized primarily by the momentum dependence rather than the detailed angular dependence of $F\left(p,\mu\right)$. Accordingly, the upper triangle ($j\ge i$) of $M_{ij}$ with even $i$ and $j$ suffices for those purposes: it determines the isotropic part of $F\left(p,\mu\right)$ and the leading anisotropic components required for macroscopic plasma quantities such as the pressure tensor. Note that the current carried by nonrelativistic particles can also be obtained exactly (Appendix~\ref{sec:Odd-Moments}).

Finally, additional formal conditions for the equivalence between the infinite system of moments $M_{ij}$ and $F\left(p,\mu\right)$ are associated with moment convergence (see, e.g., \citep{malkov2017exact} and references therein for further discussion). The Fokker--Planck case considered in that work differs from the DKE considered here in that it addresses particle transport in configuration space rather than in momentum space. A key consequence is that adjacent columns in the moment matrix are coupled in the transport problem, allowing the entire matrix to be resolved recursively, so that questions of equivalence between the moment system and the underlying PDE (Fokker--Planck or DKE) become more central.

To conclude this section, we have converted the evolution equation~(\ref{eq:F}) into a set of infinitely many independent subsystems of ordinary differential equations, eq.~(\ref{eq:MomEq}), for the moment matrix $M_{ij}$. Each subsystem couples matrix elements with a fixed $j$, i.e., within a single column. We have also identified the exactly closable part of each subsystem, which, for any given $j$, consists of a finite number of equations with the same number of unknown moments $M_{ij}$.

\subsection{Plasma Heating\protect\label{subsec:Plasma-Heating}}

Much about plasma energization can be learned already by closing the moment system in eq.~(\ref{eq:MomEq}) at $j=2$. We consider the most general case of $\sigma\left(t\right)$ in eq.~(\ref{eq:3Sigmas}), corresponding to conservation of the number of particles, $N=\mathrm{const}$. This choice leads to the double-adiabatic (CGL) equation of state, $P_{\perp}\propto nB$ and $P_{\parallel}\propto n^{3}/B^{2}$ \citep{CGL56}; see also, e.g., \citep{Lichko2017}. The other two cases listed in eq.~(\ref{eq:3Sigmas}) are recovered as special cases of the $N=\mathrm{const}$ choice, namely $n\propto B$ ($\sigma=0$), as in, e.g., effectively two-dimensional motion, and $n=\mathrm{const}$ ($\sigma=2$). Using the results of the preceding section together with eq.~(\ref{eq:MomEq}), we obtain the following closed system for the moments $M_{02}$ and $M_{22}$, associated with the isotropic part of the particle pressure and the quadrupole anisotropy (quadratic in $\mu$, i.e., the $P_{2}\left(\mu\right)$ component in a Legendre decomposition) of the particle distribution, respectively:
\[
\frac{\partial M_{02}/B}{\partial t}+M_{22}\frac{n^{2}}{B^{4}}\frac{\partial}{\partial t}\frac{B^{3}}{n^{2}}=0
\]
\[
\frac{\partial}{\partial t}\left(M_{22}\frac{B^{2}}{n^{2}}e^{6t}\right)=2e^{6t}\frac{B^{2}}{n^{2}}M_{02}.
\]
Introducing the notation
\begin{equation}
u\equiv\frac{M_{02}}{B};\,\,\,v\equiv\frac{M_{02}-3M_{22}}{B};\,\,\,\beta=\frac{n^{2}}{B^{3}};\,\,\,\eta=\frac{\dot{\beta}}{3\beta},\label{eq:not-uv-beta-eta}
\end{equation}
these equations take the compact form
\begin{equation}
\dot{u}=\eta\left(u-v\right);\qquad\dot{v}+v\left(6-2\eta\right)=-2\eta u.\label{eq:SystFor-u-v}
\end{equation}
By definition of the moments $M_{02}$ and $M_{22}$, the variables $u$ and $v$ represent, respectively, the isotropic and anisotropic pressure components normalized by $B$. The two linear equations for $u$ and $v$ can be reduced to a single first-order equation for the ratio $w=v/u$, which is nonlinear (Riccati form):
\begin{equation}
\dot{w}+w\left(6-\eta\right)+2\eta=\eta w^{2}.\label{eq:Riccati}
\end{equation}
Here $w\equiv v/u$ and $\dot{u}/u=\eta\left(1-w\right)$. Since eq.~(\ref{eq:Riccati}) remains an exact consequence of eq.~(\ref{eq:F}), the driver $\eta\left(t\right)$ may be arbitrarily large. Before discussing solutions of eqs.~(\ref{eq:SystFor-u-v}) and (\ref{eq:Riccati}), we turn in the next subsection to the physical motivation for choosing the MP driver in the form $\eta=\dot{\beta}/3\beta$. Note that MP is essentially driven by the ratio $n^{2}/B^{3}\equiv\beta$, which can be multiplied by an arbitrary constant without changing the MP equations.

\subsubsection{The Choice of the MP Driver}

We have not imposed constraints on the MP driver other than its slow variation over one gyroperiod. Thus, a broad range of physically meaningful choices is available. According to the lowest-order MP description, eqs.~(\ref{eq:SystFor-u-v}) or (\ref{eq:Riccati}), the pumping enters through temporal variations of the double-adiabaticity parameter $\beta\left(t\right)\equiv n^{2}/B^{3}$. When the double-adiabatic (CGL) equation of state applies, this quantity is proportional to the pressure ratio, $\beta\propto P_{\perp}/P_{\parallel}$ \citep{CGL56}. Since eqs.~(\ref{eq:SystFor-u-v})--(\ref{eq:Riccati}) are invariant under multiplication of $\beta$ by an arbitrary constant, we fix the normalization by imposing $\overline{\beta\left(t\right)}=1$, where the overbar denotes averaging over one driver period. Accordingly, $n$ and $B$ may be nondimensionalized such that this normalization is satisfied. We therefore specify the driver in the form
\begin{equation}
\beta\left(t\right)\equiv\frac{n^{2}}{B^{3}}=1+A\sum_{k=1}^{n}a_{k}\sin\left(kVt+\alpha_{k}\right)\label{eq:BetaOf-t}
\end{equation}
The effective driver strength $\eta$ in eq.~(\ref{eq:Riccati}) is then controlled by the dimensionless amplitude parameter $A$, the number of modes, and their relative weights $a_{k}$, which we parameterize by a spectral index $q$ via $a_{k}^{2}=k^{-q}$. The frequency $kV$ of the $k$th mode contains a dimensionless parameter $V$, which may be related, for example, to a magnetosonic propagation speed. In that case, correlated $B$ and $n$ perturbations can be produced by an ensemble of long-wavelength magnetosonic modes after averaging over a magnetic flux tube, as discussed in Sec.~\ref{subsec:Heating-and-Accel}.

To narrow the choice of $\beta\equiv n^{2}/B^{3}$, we first consider representative relations between turbulent density and magnetic-field variations in astrophysical settings. Because the $n$--$B$ relation is a subject of active interest across several communities, particularly those studying star formation and ISM turbulence, we begin with their salient observational constraints. From an analysis of Zeeman measurements in 27 molecular clouds, \citet{CrutcherMC99} reported a scaling $\left|\boldsymbol{B}\right|\propto n^{\kappa}$ with $\kappa=0.47\pm0.08$ over the density range $10^{3}$--$10^{4}\,\mathrm{cm}^{-3}$. This is consistent with the physically motivated scaling $\left|B\right|\propto n^{1/2}$ (also discussed in \citealp{CrutcherMC99}), corresponding to an approximately constant Alfv\'en speed. While subsequent work indicates that the effective $B$--$n$ scaling can depend on environment and may exhibit breaks between different regimes, a $B\propto n^{1/2}$ scaling is among the commonly discussed possibilities (see, e.g., the compilation and discussion in \citealp{Whitworth2024}). Under $B\propto n^{1/2}$, one obtains $\beta\propto n^{2}/B^{3}\propto B$, so that $\beta$ varies in phase with $B$ and drives MP. It is also instructive to note that, for effectively two-dimensional motion across $\boldsymbol{B}$, one expects $B\propto n$, implying $\beta\propto B^{-1}$; in this case the MP driver $\eta=\dot{\beta}/3\beta$ changes sign, while the formal structure of the closed moment system remains unchanged. Finally, for the intermediate scaling $B\propto n^{2/3}$ one has $\beta=\mathrm{const}$, so that MP is not driven at this (lowest-order) level.

\begin{figure}
\includegraphics[viewport=140bp 0bp 616bp 382bp,scale=0.3]{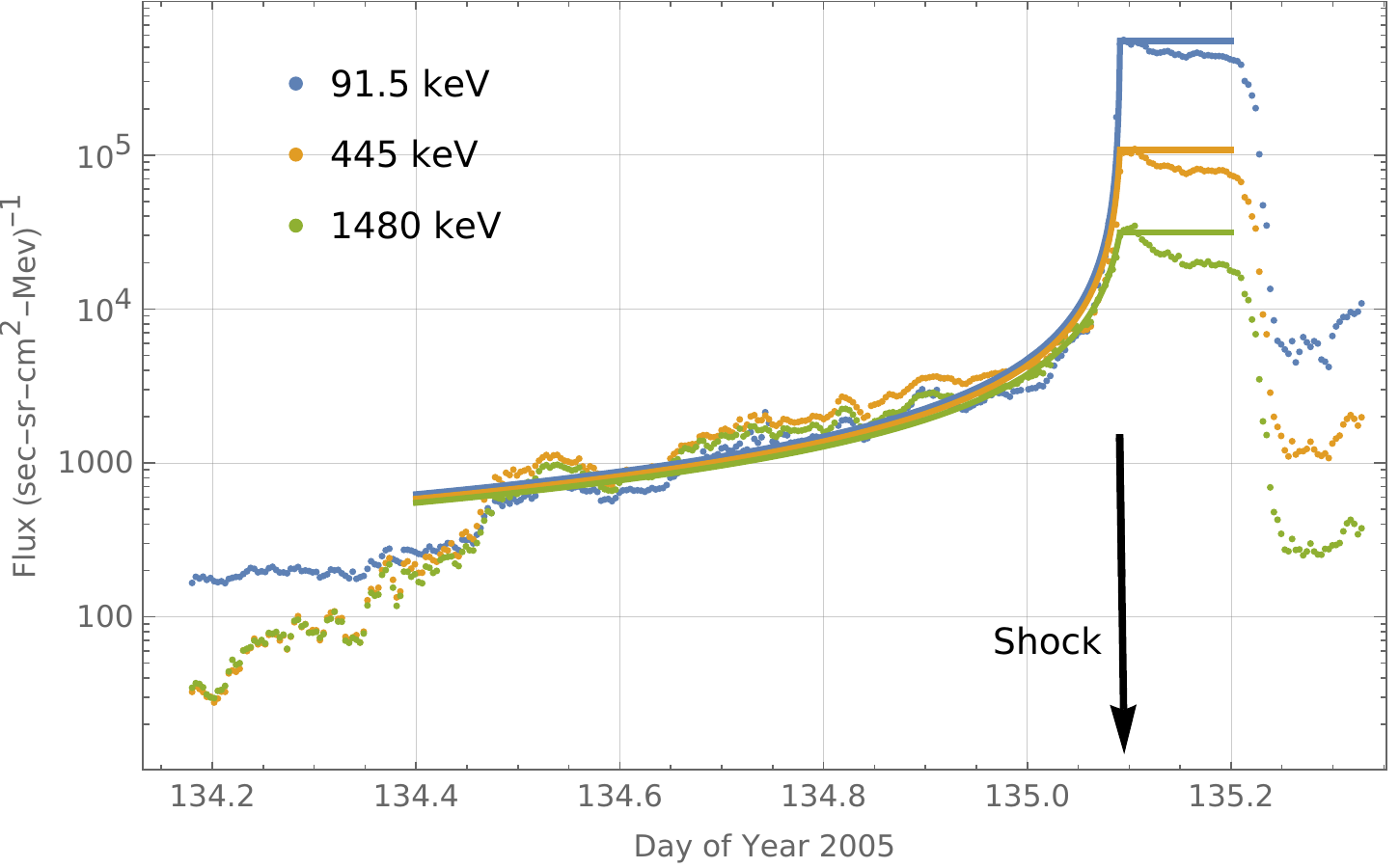}
\includegraphics[scale=0.45]{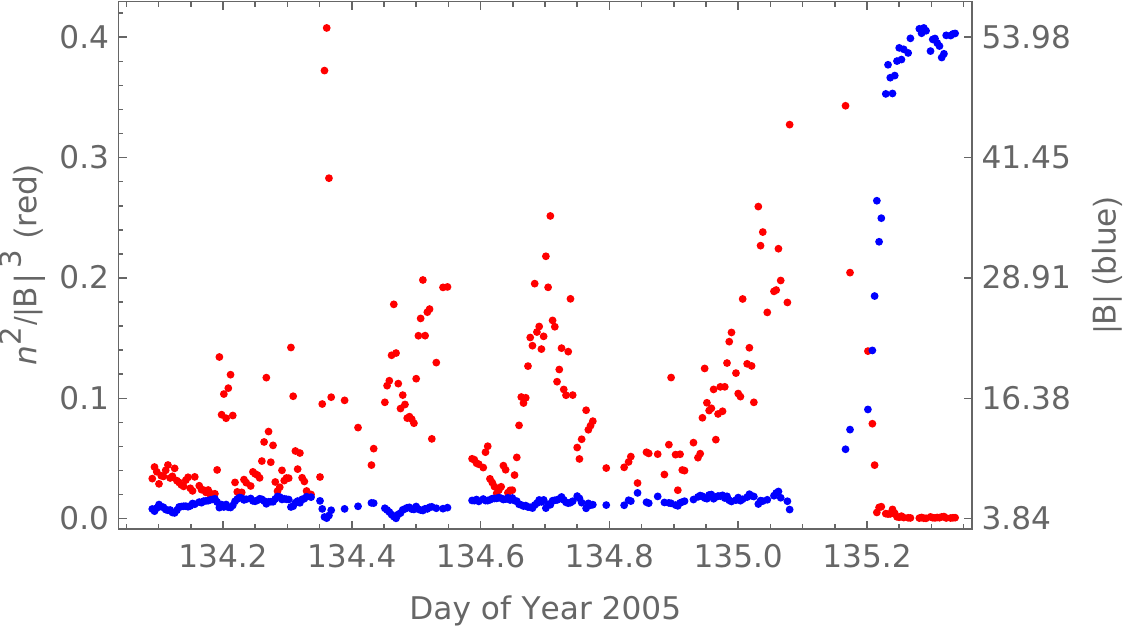}
\caption{March 2005 interplanetary shock: \textbf{Top Panel:} Particle spectra upstream (left to the shock) and downstream (on its right side). \textbf{Bottom Panel:} Magnetic field magnitude, $B$, and plasma density. \protect\label{fig:March-2005-interplanetary}}
\end{figure}

Meanwhile, \citet{Whitworth2024} provide detailed $B$--$n$ scalings across different ISM conditions using multiple observational analyses and simulation approaches. A common feature is the presence of breaks between distinct $B$--$n$ power-law regimes, with the break position depending on density. More specific, effectively point-by-point measurements of $B$ and $n$ are available from \emph{in situ} heliospheric data. Of particular interest here are spacecraft measurements obtained during shock crossings, where both $B$ and $n$ can vary strongly and coherently on time scales comparable to the characteristic scattering time. We provide specific examples below that may be indicative of a localized MP-active region in the shock environment. 

\begin{figure}
\includegraphics[scale=0.35]{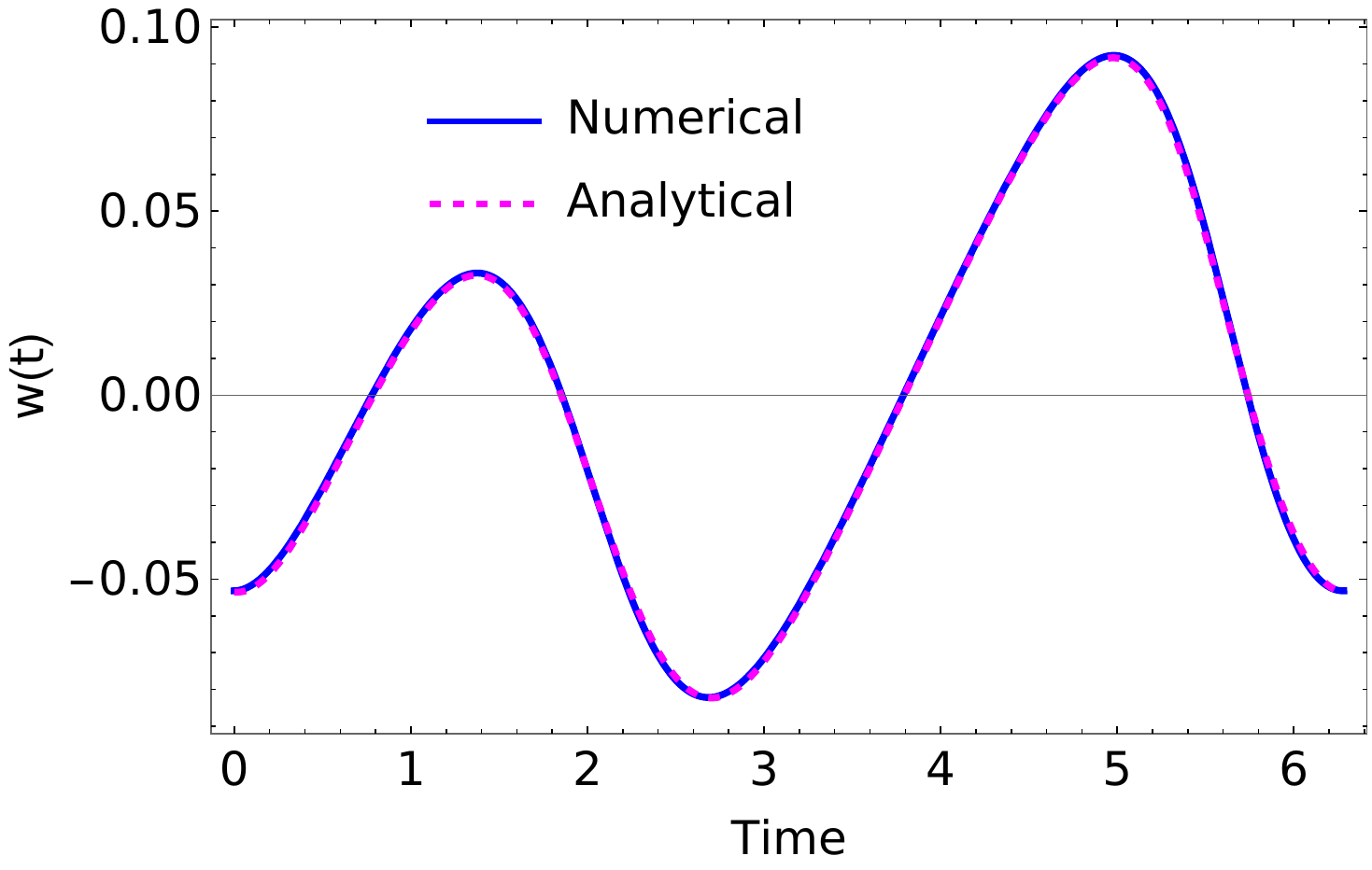}
\includegraphics[scale=0.36]{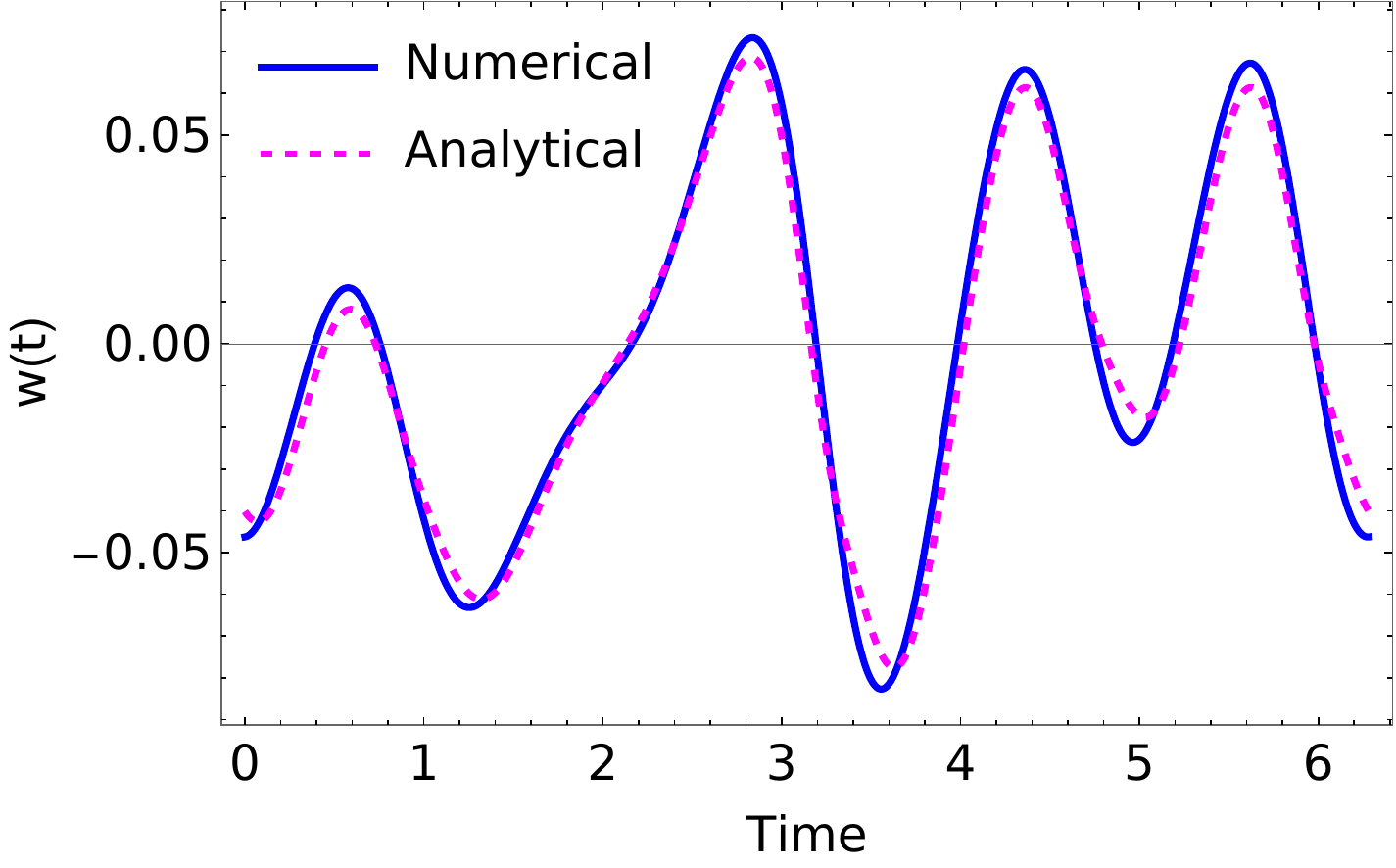}
\caption{Comparison of analytic solution of eq.(\ref{eq:Riccati}) presented in eq.(\ref{eq:w-corr}) with the numerical solution of the same equation shown for two different numbers of the driver's harmonics, but similar total power with the same spectral index, $q=1$. This setup is consistent with an isotropic, resonant pithch-angle particle scattering. \textbf{Top panel}: $A\approx0.41$, $n=2$; \textbf{Bottom panel}: $A\approx0.17$, $n=5$. \protect\label{fig:Compar-analytic-num-w-1}}
\end{figure}

\subsubsection{Some Common Approximations to MP and their Limitations}

We have reduced the mathematical description of MP from the general DKE to a pair of linear, multiplicatively driven (by the driver $\eta\left(t\right)$) first-order ordinary differential equations (ODEs), eqs.~(\ref{eq:SystFor-u-v}), without making additional approximations. Equivalently, the same dynamics can be expressed in terms of a single first-order but nonlinear (Riccati) ODE, eq.~(\ref{eq:Riccati}). There are multiple ways to analyze these equations. Here we employ two complementary approaches that are common in the MP literature: an adiabatic approximation based on a slowly varying driver, and a perturbative (quasilinear) approximation based on a small-amplitude driver. After assessing their limitations, we return to an exact, but numerically assisted, treatment. The adiabatic approximation assumes that $\eta\left(t\right)$ varies slowly compared to the pitch-angle relaxation time. While it can treat strong drivers provided they vary slowly, it breaks down when $\eta$ contains a fast-time component, even if that component has relatively small amplitude, and is therefore not suitable for broadband magnetic turbulence; many MP models, in practice, are restricted to single-mode or narrowband driving. The perturbative approach, by contrast, assumes a low-amplitude driver and is also often implemented in a single-mode regime (e.g., \citealp{Lichko2017}); however, when combined with a random-phase approximation it can be applied to an essentially arbitrary bandwidth, and is therefore complementary to the adiabatic approach.

\paragraph{Adiabatic Approximation}

We begin with moderate values of $\eta$, assuming also that its variation is slow in the sense that $\eta/\dot{\eta}>1$. In this case, an adiabatic approximation is appropriate. To zeroth order, we drop the time derivative in the Riccati equation, eq.~(\ref{eq:Riccati}), and obtain
\begin{equation}
w_{0}\left(t\right)=-\frac{2\eta(t)}{\sqrt{2\eta\left(t\right){}^{2}+\left[3-\eta(t)/2\right]{}^{2}}+3-\eta(t)/2}\label{eq:w0}
\end{equation}
The utility of this approximation follows from the relatively short linear relaxation time $\tau_{\text{rel}}=1/6$, so that for $\eta\ll1$ it reduces to $w_{0}\simeq-\eta/3$. Recall that time is measured in units of the characteristic isotropization time. Another reason to use this form is that $\eta$ is assumed periodic, so $w_{0}$ must also become periodic, at least in the time-asymptotic regime. This is not generally true for the variables $u$ and $v$, each of which can grow exponentially in time according to Floquet theory for linear differential equations with periodic coefficients (e.g., \citealp{Coddington1955}). In particular, each variable can be represented as $p\left(t\right)\exp\left[\gamma t\right]$, where $p\left(t\right)$ is a $\tau$-periodic function and $\gamma$ is a constant. We return to this point below when analyzing the basic system in eqs.~(\ref{eq:SystFor-u-v}). Once a periodic asymptotic regime for the solution of eq.~(\ref{eq:Riccati}) is established, $u$, $v$, and the growth rate $\gamma$ follow from
\begin{equation}
\ln u=\frac{1}{3}\ln\beta-\int\eta w\,dt,\qquad v=wu,\qquad\gamma\equiv-\frac{1}{\tau}\int_{0}^{\tau}\eta w\,dt\label{eq:GrRate-gamma}
\end{equation}
We can improve upon the approximation $w_{0}\left(t\right)$ in eq.~(\ref{eq:w0}) by including the correction associated with $\dot{w}$ in eq.~(\ref{eq:Riccati}). The correction must be constructed so as to eliminate secular terms. A convenient way to achieve this is to incorporate the correction through a time shift, $t\to t-\Delta t$, where $\Delta t$ depends on $\eta\left(t\right)$. The resulting approximation may be written as
\begin{eqnarray}
w\left(t\right) & \approx & w_{0}\left(t-\Delta t\right)+\left[\frac{1}{2}\ddot{w}_{0}\left(t-\Delta t\right)-\dot{w}_{0}^{2}\left(t-\Delta t\right)\right]\Delta t^{2}\nonumber \\
 & + & 2\dot{w}_{0}\left(t-\Delta t\right)\chi\left(t-\Delta t\right)\Delta t,\label{eq:w-corr}
\end{eqnarray}
where we use the notation
\begin{equation}
\Delta t\left(\eta\right)\equiv\frac{1}{6-\eta\left(t\right)+2\eta^{2}\left(t\right)/3},\qquad\chi\left(t\right)\equiv\frac{d\Delta t}{dt}.\label{eq:DefsEtaDelta-t}
\end{equation}
To test the accuracy and applicability of this approximation, we substitute the MP driver $\beta\left(t\right)$ defined in eq.~(\ref{eq:BetaOf-t}) into the approximation in eq.~(\ref{eq:w-corr}). We use two representative spectra for $\beta$: one containing only two harmonics and another containing five harmonics, both with spectral index $q=1$.

For a slowly varying $\eta\left(t\right)$ containing only two harmonics, the approximation in eq.~(\ref{eq:w-corr}) agrees well with direct numerical integration of the Riccati equation, eq.~(\ref{eq:Riccati}). As expected from the construction of the adiabatic approximation, its accuracy deteriorates when higher-frequency components are added to the driver $\eta$, as in the five-harmonic example. We illustrate this behavior in Fig.~\ref{fig:Compar-analytic-num-w-1} by comparing analytic and numerical solutions of eq.~(\ref{eq:Riccati}) for representative slow and relatively fast driver variations.

The accuracy of the asymptotic expansion in eq.~(\ref{eq:w-corr}) depends on both the magnitude and time variability of the driver $\eta\left(t\right)=\dot{\beta}/3\beta$ in eq.~(\ref{eq:Riccati}). It is most accurate for sufficiently steep spectra with a limited number of modes $n$. We have plotted the approximation in eq.~(\ref{eq:w-corr}) in Fig.~\ref{fig:Compar-analytic-num-w-1} against a direct numerical integration of eq.~(\ref{eq:Riccati}). The approximation becomes progressively less accurate as the number of modes increases, in which case a different analytic strategy based on the alternative form of the basic system, eq.~(\ref{eq:SystFor-u-v}), is more appropriate. Perhaps the most useful consequence of the simplest adiabatic estimate, $w_{0}\simeq-\eta/3$, is the explicit expression it yields for the growth rate in eq.~(\ref{eq:GrRate-gamma}),
\[
\gamma\equiv\frac{1}{3\tau}\int_{0}^{\tau}\eta^{2}\,dt.
\]
With a growing number of modes and, to a lesser extent, with increasing driver amplitude, the adiabatic approximation becomes progressively inaccurate. We therefore turn next to the standard quasilinear treatment appropriate to broadband driving with random phases.

\paragraph{Quasilinear/Random-Phase Approximation}

It is convenient to use an alternative form of the basic equations in eq.~(\ref{eq:SystFor-u-v}), namely a single second-order equation for the dependent variable
\[
\psi=\left(u-v\right)/\beta,
\]
which obeys
\begin{equation}
\left(\beta\psi^{\prime}\right)^{\prime}+6\beta\psi^{\prime}+4\beta^{\prime}\psi=0\label{eq:psi-sec-ord-eq}
\end{equation}
Here prime denotes the time derivative. The variables $u$ and $v$ can be recovered from $\psi$ using the above definition together with $\beta\psi^{\prime}=6v$. To shorten the notation, we rewrite the driver representation in eq.~(\ref{eq:BetaOf-t}) in the form
\begin{equation}
\beta=1+\sum_{k=-\infty}^{\infty}\beta_{k}e^{i\omega_{k}t}\equiv1+\tilde{\beta}\left(t\right)\label{eq:betaSer}
\end{equation}
where we have replaced $kV$ with $\omega_{k}$ to make the derivation more general, and we assume that the complex amplitudes $\beta_{k}$ have random phases and satisfy $\beta_{-k}=\beta_{k}^{*}$ so that $\beta$ remains real. Here the star denotes complex conjugation and $\beta_{0}\equiv0$. It is then natural to seek $\psi\left(t\right)$ in a similar series,
\begin{equation}
\psi=\sum_{k=-\infty}^{\infty}\psi_{k}e^{i\omega_{k}t}\label{eq:psiSer}
\end{equation}
Our main interest is in the slowly varying component $\psi_{0}\left(t\right)$, corresponding to $\omega_{0}=0$, which we assume does not carry a random phase. Substituting the series (\ref{eq:betaSer})--(\ref{eq:psiSer}) into eq.~(\ref{eq:psi-sec-ord-eq}) and averaging over the random phases of the ensemble $\left\{\beta_{k}\right\}$, we obtain (details are given in Appendix~\ref{sec:Random-Phase-Approximation})
\begin{equation}
\psi_{0}^{\prime}=\frac{8}{3}\psi_{0}\sum_{k=1}^{\infty}\frac{\omega_{k}^{2}\left|\beta_{k}\right|^{2}}{36+\omega_{k}^{2}}\equiv\gamma\psi_{0}\label{eq:psi0prime}
\end{equation}
The lowest harmonic in the spectrum $\left\{\omega_{k}\right\}$ is $\omega_{1}$, so the driver period is $2\pi/\omega_{1}$. Writing $\psi_{0}^{\prime}=\gamma\psi_{0}$ and denoting the Floquet multiplier by $\rho=\psi_{0}\left(t+2\pi/\omega_{1}\right)/\psi_{0}\left(t\right)=\exp\left(2\pi\gamma/\omega_{1}\right)$, we obtain
\begin{equation}
\rho=\exp\left(\frac{16\pi}{3\omega_{1}}\sum_{k=1}^{\infty}\frac{\omega_{k}^{2}\left|\beta_{k}\right|^{2}}{36+\omega_{k}^{2}}\right)\label{eq:rho-QL}
\end{equation}

\begin{figure}
\includegraphics[viewport=0bp 0bp 540bp 370bp,scale=0.45]{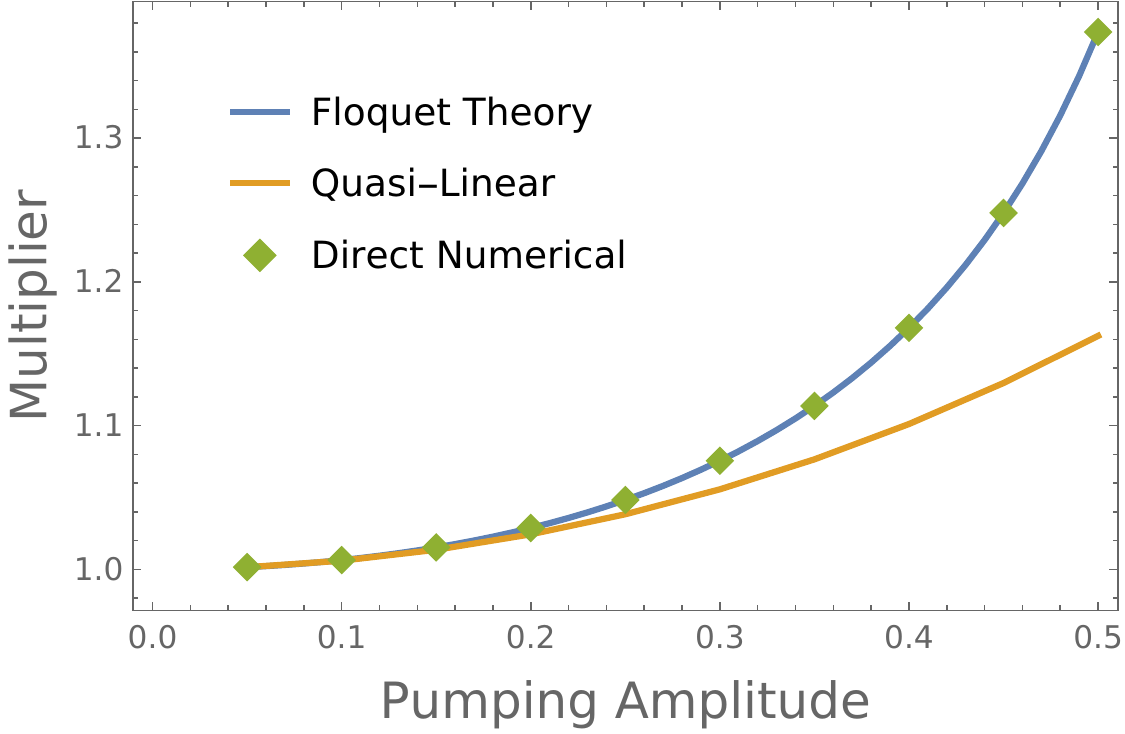}
\caption{Comparison of three different approaches to the solution of eqs.(\ref{eq:SystFor-u-v}): The Floquet multiplier $\rho_{1}$, eq.(\ref{eq:FloquetMult}) is shown alongside with the direct numerical integration of eq.(\ref{eq:psi-sec-ord-eq}) (see text). \protect\label{fig:FTsingleDriver}}
\end{figure}

We can verify this result directly by solving eq.~(\ref{eq:psi-sec-ord-eq}) numerically and evaluating the growth of $\psi_{0}$ over one driver period. 
For this example, we replace A in eq.([eq:betaSer-1]) by its normalized value $a=A\left(\sum_{k=\text{1}}^{n}a_{k}^{2}\right)^{-1/2}$ to represent $\beta$  as 

\begin{equation}
\beta=1+a\sum_{k=\text{1}}^{n}k^{-q/2}\sin\left(kVt+\alpha_{k}\right)\label{eq:betaSer-1}
\end{equation}
with $n=3$ and $V=1$, vary the amplitude over $0.05<a<0.5$, and set $q=1/2$, which is often expected in shock environments where MHD turbulence self-generated by accelerated particles may approach a Bohm-like regime with spectral density of magnetic fluctuations $\propto1/k$. The Floquet multiplier is shown in Fig.~\ref{fig:FTsingleDriver} as a function of the driver amplitude $a$ for $n=3$. In addition to the numerical solution, the Floquet-theory result is also shown for comparison, to which we turn next.

\begin{figure}
\includegraphics[scale=0.45]{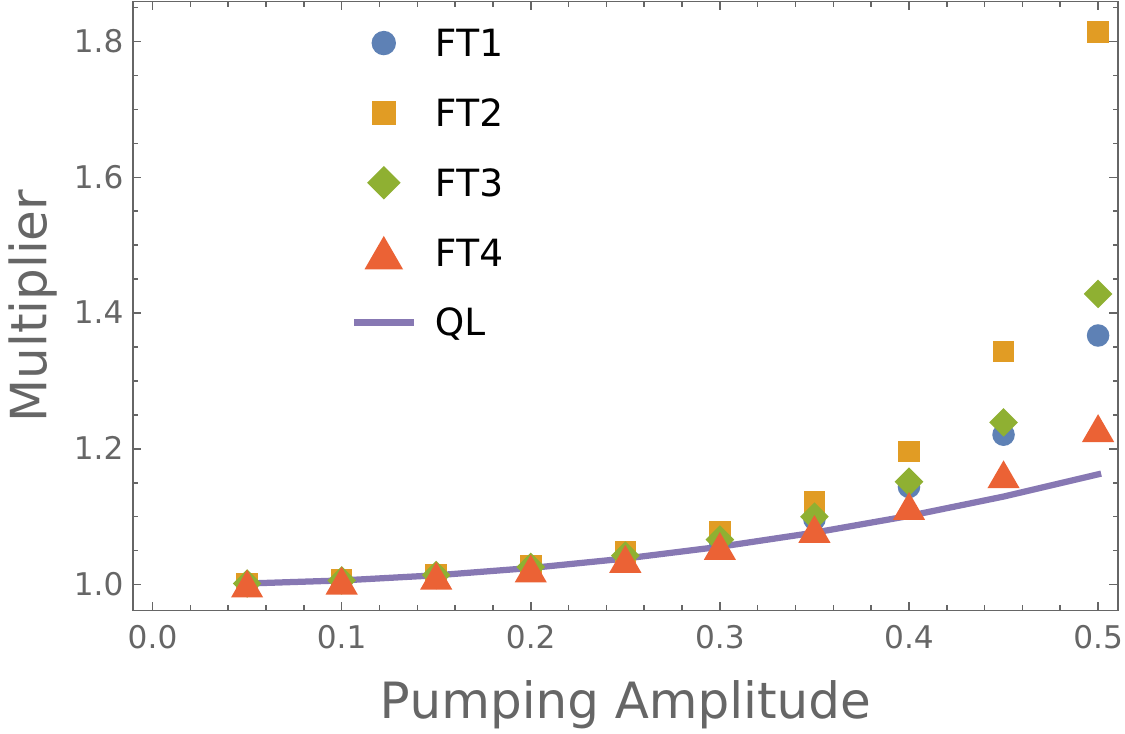}
\caption{Same as in Fig.\ref{fig:FTsingleDriver} except for the results of the Floquet's theory (FT 1-4) calculated for four different realizations of random phases and omitted direct numerical solution as it was shown to be identical to the FT results.\protect\label{fig:FTrandomEnsemble}}
\end{figure}

\subsubsection{Exact Floquet Multipliers\protect\label{subsec:Exact-Floquet-Multiplier}}

As is seen from Fig.~\ref{fig:FTsingleDriver}, the perturbative approach discussed above becomes increasingly inaccurate once the driver amplitude reaches $a\simeq0.2$--$0.3$. We therefore use the MP equation in the form of eq.~(\ref{eq:psi-sec-ord-eq}). Once a pair of linearly independent solutions of this equation is obtained numerically over one period of the driver, say $0\le t<\tau$, Floquet theory provides the solution for all $t>0$. Thus, it suffices to compute two linearly independent solutions over a single driver period.

As shown in Appendix~\ref{sec:Floquet-Calculations}, if we start at $t=0$ with $\psi_{1}\left(0\right)=1$, $\psi_{1}^{\prime}\left(0\right)=0$ and $\psi_{2}\left(0\right)=0$, $\psi_{2}^{\prime}\left(0\right)=1$, then the Floquet multipliers can be calculated as
\begin{equation}
\rho_{1,2}\text{=}\frac{1}{2}\left(\psi_{2}^{\prime}+\psi_{1}\right)\pm\sqrt{\frac{1}{4}\left(\text{\ensuremath{\psi_{2}^{\prime}}}+\psi_{1}\right){}^{2}-\beta\left(0\right)e^{-6\tau}}\label{eq:FloquetMult}
\end{equation}
Here $\psi_{1,2}$ denote the two linearly independent solutions evaluated at $t=\tau$ with the above initial conditions imposed at $t=0$. The faster growing component corresponds to $\rho_{1}$, which is shown in Fig.~\ref{fig:FTsingleDriver}. Time asymptotically, almost any solution of eq.~(\ref{eq:psi-sec-ord-eq}) grows at the same rate, $\psi\left(t+\tau\right)=\rho_{1}\psi\left(t\right)$. Also shown in Fig.~\ref{fig:FTsingleDriver} is the direct numerical solution of eq.~(\ref{eq:psi-sec-ord-eq}) in the form of the ratio $\psi\left(2\tau\right)/\psi\left(\tau\right)\approx\rho_{1}$.

Our interest is primarily in the Floquet multiplier $\rho_{1}$, since it represents the faster growing component of the solution of eq.~(\ref{eq:psi-sec-ord-eq}). In particular, $\rho_{1}^{2}-\rho_{2}^{2}=\beta\left(0\right)e^{-6\tau}>0$. In writing eq.~(\ref{eq:FloquetMult}) we assume that the sign of $\psi_{1}\left(t\right)+\psi_{2}^{\prime}\left(t\right)$, which equals $2$ at $t=0$, does not change over the interval $0\le t\le\tau$. This is unlikely for plausible choices of $\beta\left(t\right)$, but can be addressed straightforwardly if it does. Although the $\psi_{1,2}$ basis is technically convenient for the Floquet analysis, the $u,v$ representation is physically more transparent. Indeed, according to eq.~(\ref{eq:SystFor-u-v}), the component $v\left(t\right)$ corresponds to the quadrupole (anisotropic) part of the angular distribution and is strongly damped by pitch-angle scattering, whereas $u\left(t\right)$ represents the isotropic component. Note, however, that both $u$ and $v$ can grow secularly in time through the multiplicative driving.

Next, we investigate the MP efficiency as a function of the particular realization of random phases in the driver's harmonic ensemble. This dependence is very strong, as demonstrated in Fig.~\ref{fig:FTrandomEnsemble}. Shown are four realizations of the driver phases $\left\{\alpha_{k}\right\}\in\left(0,2\pi\right)$ in the series of eq.~(\ref{eq:betaSer-1}). For comparison, the analytic result based on the quasilinear random-phase approximation, eq.~(\ref{eq:rho-QL}), is shown as a solid line. As expected, with increasing driver amplitude the heating efficiency becomes progressively more sensitive to the particular phase realization of the driving modes. This sensitivity effectively devalues the random-phase approximation when the amplitude approaches $a\simeq0.4$--$0.5$, making Floquet theory the method of choice in this regime of driver amplitudes. 

\subsection{Particle Acceleration\protect\label{subsec:Particle-Acceleration}}

A key advantage of the above approach is that it provides a controlled
route from heating to acceleration. The following two aspects of this
transition are particularly advantageous. First, as soon as the solution
is found within one period of the driver, it can be extended indefinitely
in time with no loss of accuracy, since it maps to the next period
by an exact Floquet multiplier (more generally, by a monodromy matrix). Second, the one-period solution itself
yields, in the low-order closure, an exact pressure tensor directly
from the full particle distribution. The latter is governed by the
drift-kinetic equation, but there is no need to solve it to obtain
the pressure tensor. The theory, therefore, provides its evolution
for an arbitrary driver amplitude, with the driver entering through
$\eta\left(t\right)$. At higher order, the same framework can in
principle track the evolution of higher moments that determine the
formation of nonthermal tails. This suggests a practical strategy
in broadband turbulence: use measured or simulated $\beta\left(t\right)$
statistics to quantify $\eta\left(t\right)$, apply the low-order
closure as a stringent test of whether the environment can plausibly
pump heat at the observed level, and then use higher-order closures
where needed to predict spectral slopes and maximum energies, with
scattering treated as a constrained parameter rather than an ad hoc
$\tau$-approximation.

\begin{figure}
\includegraphics[scale=0.35]{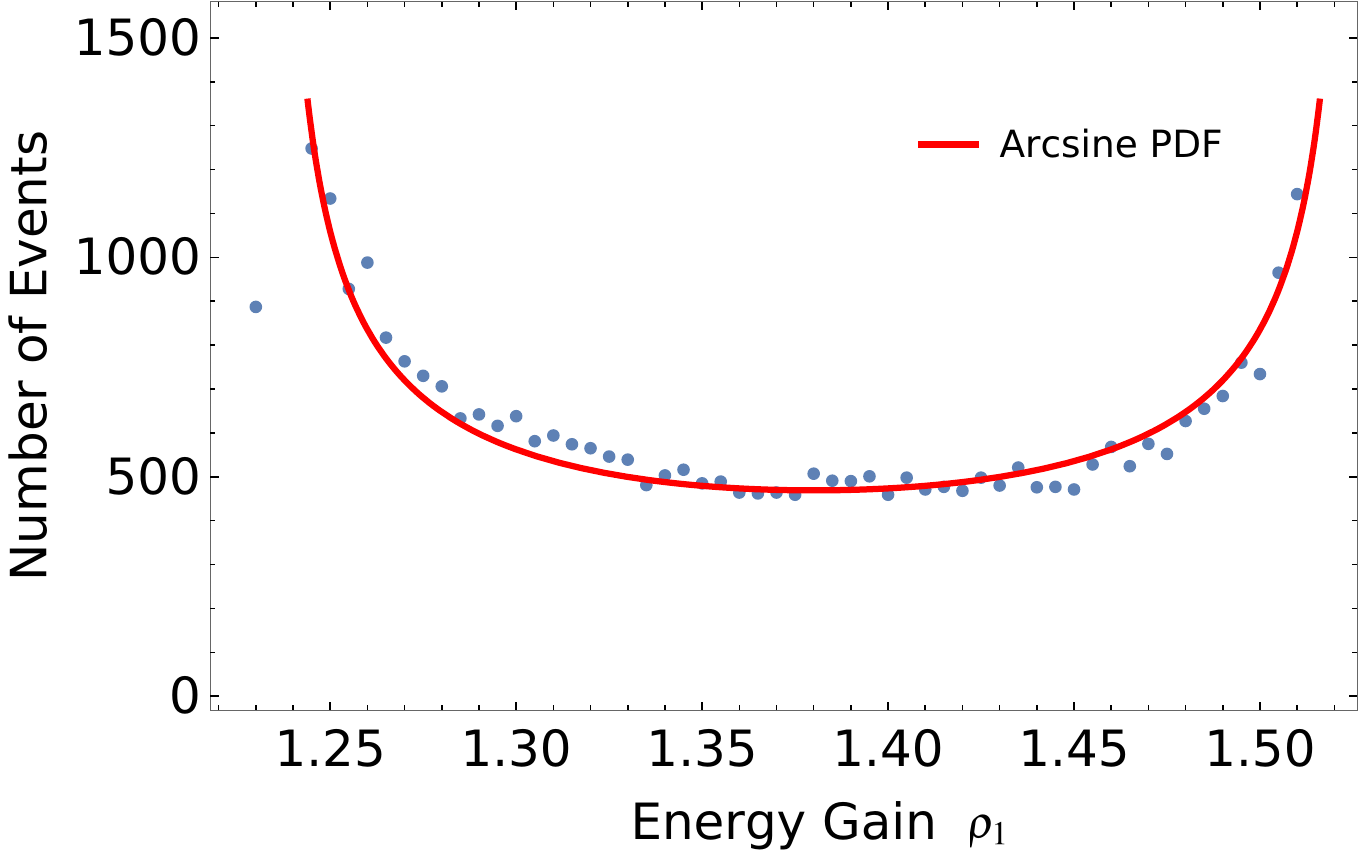}

\caption{Distribution of acceleration events for $n=2$ harmonics in the driver,
nuber of acceleration events $N=16000,$ the PL spectral parameter of the driver $q=1$,
and normalized amplitude $a=0.38$. \protect\label{fig:ArcsinePDF}}
\end{figure}

While deferring the implementation of the above strategy to a future
publication, we demonstrate below that energizing particles during
only one period of the driver can generate suprathermal power-law
tails, if multiple driver modes are at play. Indeed, when testing
our closure scheme using a limited number of harmonics, we found that
a mere broadening of the driver's bandwidth dramatically enhances
the heating efficiency for certain phase distributions of the driver
modes, as exemplified in Fig.\ref{fig:FTrandomEnsemble} where only three
driver modes are active. It is, therefore, natural to start our investigation
of particle acceleration by tracking energy gain vs mode number. 

Time asymptotically, phase randomization of a single mode does not
have a measurable impact on the heating rate. By contrast, already
two random modes energize particles much more efficiently, and their
phase difference critically affects the energy gain. We, therefore,
begin with only two modes, whose phases are randomly changed. All
other driver parameters remain the same. Using this setting, we compute
a relatively large number of one-period solutions to the MP equation
to generate a statistically sufficient distribution of Floquet multiplier
(i.e., heating efficiency). The result of this simple numerical
experiment is presented in Fig.\ref{fig:ArcsinePDF}, which shows
the distribution for the case of two modes. The familiar Arcsine distribution
fit is shown for comparison. It clearly points to the bimodal distribution
of the heating rate in this regime. A noticeable skewness is present,
but we have not researched further into its statistical significance. 

\begin{figure}
\includegraphics[scale=0.35]{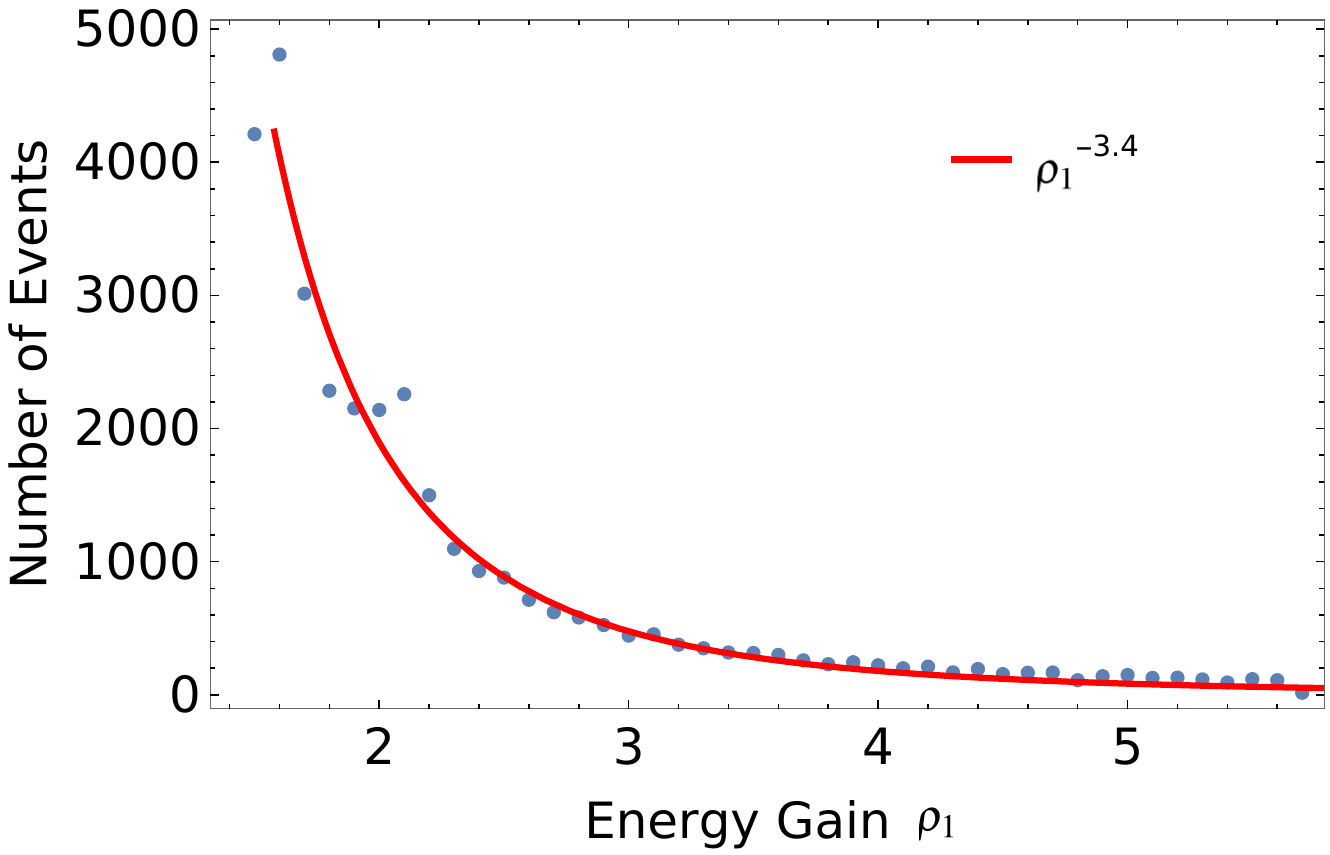}

\caption{Same as in Fig.\ref{fig:ArcsinePDF} but for the number of modes $n=3$, normalized
amplitude $a=0.38$.\protect\label{fig:mode3-SmalEns}}
\end{figure}

\begin{figure}
\includegraphics[scale=0.35]{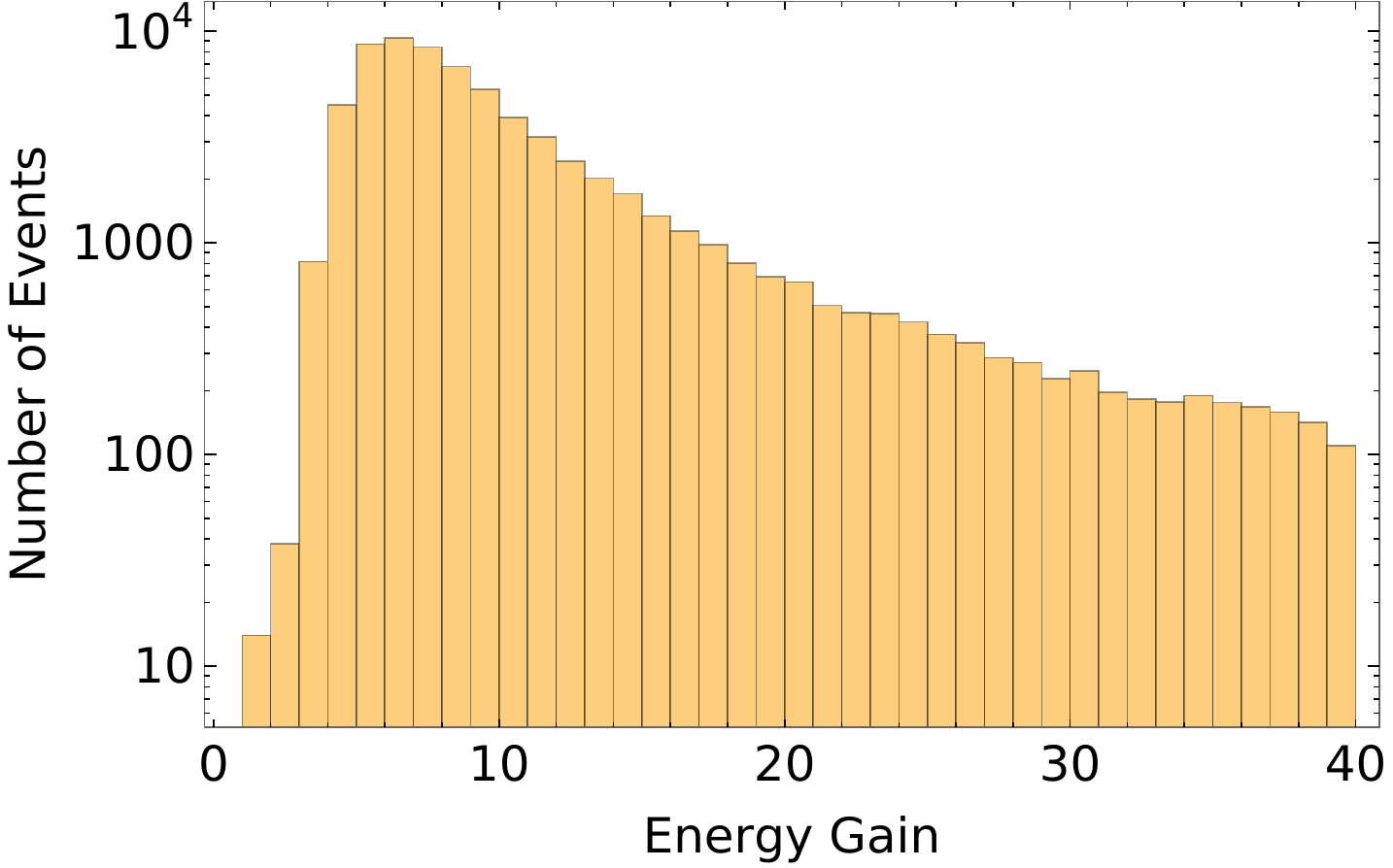}\caption{Histogram of the Floquet inexes for $n=10$, $a=0.27$. \protect\label{fig:HistogrOfEvents-n10a027} }
\end{figure}

\begin{figure}
\includegraphics[scale=0.3]{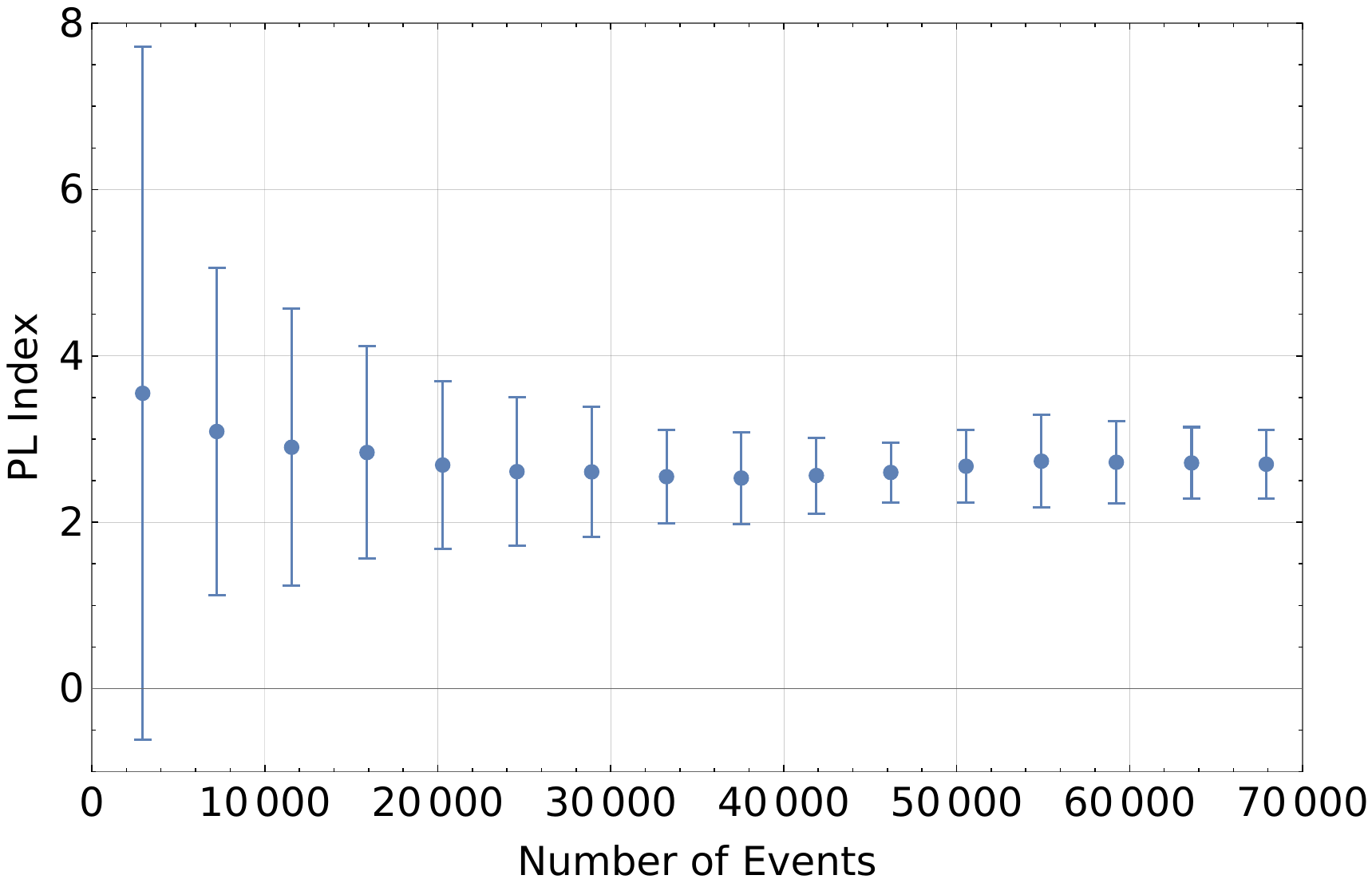}

\caption{Power law index of the tail of the distribution shown in Fig.\ref{fig:HistogrOfEvents-n10a027}.
\protect\label{fig:Power-law-index}}
\end{figure}

Next, we have increased the number of modes by 1, setting $n=3$ in
the driver spectrum. The bimodal Arcsine type distribution of the
Floquet multiplier turned into an almost regular power law, as shown
in Fig.\ref{fig:mode3-SmalEns}. However, a small remnant of the higher
end of the Arcsine distribution for $n=2$ appears to be present around
$\rho\simeq2$. To further investigate its behavior, we continued
increasing the number of modes while also enhancing the statistics.
We found that the distribution becomes smoother and the power-law
index appears to relax to lower values, compared to the $\approx3.4$
shown in Fig.\ref{fig:mode3-SmalEns}. Not surprisingly, it depends
on the driver amplitude. Also, with the growing number of modes, the
maximum energy gain in the tail of the distribution increases. The
respective results are shown in Figs.\ref{fig:HistogrOfEvents-n10a027}
and \ref{fig:Power-law-index}

\section{Discussion\protect\label{sec:Discussion}}

In this paper we have approached the magnetic pumping particle energization
process using the method of moments. Contrasting to most of the other
approaches, we pursued an exact closure of an infinite system of ODEs
for the moments resulting from the original PDE's conversion. Since
the exact closure does not require a small parameter, which normally
constitutes the pumping amplitude, both heating and subsequent particle
acceleration were addressed within the suggested framework. 

The usefulness of the exact moment closure can be appreciated more
fully by unifying descriptions of the particle transport in the momentum
space, which was the subject of the present study, with the parallel
and perpendicular particle transport to the magnetic field in coordinate
space. The overarching aspect for this unification is the same form
of the pitch-angle scattering operator (Lorentzian gas) in all three
cases. The pitfalls of the truncation method are better understood
in the spatial transport problems and have been discussed with regard
to the parallel particle diffusion in \citep{malkov2017exact} while
an exact form of the recently obtained perpendicular diffusion coefficients
speaks for itself by its strong variability at the initial propagation
stage \citep{MalkovLemoine2023}.

\section*{ACKNOWLEDGMENTS}

M. M. was supported by NASA ATP 80NSSC24K0774 and Fermi 80NSSC25K7346
awards. I.C.J. acknowledges support from the Research Council of Finland
(X-Scale, grant No. 371569), and from ISSI\textquoteright s \textquotedblleft Visiting
Scientist Program\textquotedblright . 

\appendix

\section{Odd Moments\protect\label{sec:Odd-Moments}}

In this Appendix we provide some justification of focusing on the
even moments $M_{2k,2n}$ with where $k,n=0,1,...$ . For, we consider
the MP driving requirements for the odd moments $M_{2k+1,2n+1}$ to
be amplified within the system of moments in eq.(\ref{eq:MomEq}).
Substituting $k=n=0$ we arrive at the first closed system of this
category that comprises only one equation for $M_{11}$. It can be
manipulated into the following form:

\begin{equation}
\frac{\partial}{\partial t}\left(\frac{B}{n}M_{11}\right)+2\frac{B}{n}M_{11}=0.\label{eq:M11}
\end{equation}
Assuming that $B$ and $n$ are $2\pi$ -periodic function of time,
we confirm the unconditional damping of the moment $M_{11}$:

\begin{equation}
M_{11}\left(t\right)=\left.M\frac{B}{n}\right|_{t=0}\frac{n\left(t\right)}{B\left(t\right)}e^{-2t}\label{eq:M11-damping}
\end{equation}
Note, that for nonrelativistic particles this moment is responsible
for the current along the magnetic field, which vanishes asymptotically
by pitch-angle scattering and is not supported by MP. 

The next subsystem of eq.(\ref{eq:MomEq}) can be closed at the level
$M_{33}$. It couples the moments $M_{13}$ with $M_{33}$ and can
be written in a compact form as follows:
\begin{equation}
\dot{\phi}+2\phi=\dot{\beta}\psi;\qquad\dot{\psi}+12\psi=6\beta^{-1}\phi.\label{eq:M13-clos}
\end{equation}
Here we used the notation:
\[
\phi=\frac{M_{13}}{n};\qquad\psi=\frac{B^{3}M_{33}}{n^{3}};\qquad\beta=\frac{n^{2}}{B^{3}}.
\]
Similarly to eqs.(\ref{eq:SystFor-u-v}), the above system contains
two unknown moments, but the difference is that both moments are strongly
damped, whereas the moment $M_{02}$ in eqs.(\ref{eq:SystFor-u-v})
is not. This remark notwithstanding, the excitation of the normalized
moments $\phi$ and $\psi$ is not impossible if the multiplicative
driver $\beta\left(t\right)$ is sufficiently strong. Nevertheless,
below we provide a condition under which this excitation is ruled
out and the MP heating and MP particle acceleration can be described
by the even moments $M_{02}$, $M_{04},\dots$, which we have focused
on throughout this paper. The condition of asymptotic decay of both
$\phi$ and $\psi$ in eq.(\ref{eq:M13-clos}) can be formulated by
introducing the following notation

\[
\gamma=\dot{\beta}+6\beta^{-1};\qquad\psi^{\prime}=\sqrt{6}\psi.
\]
From eqs.(\ref{eq:M13-clos}) we obtain

\begin{equation}
\frac{1}{2}\left(\left\Vert \phi\right\Vert ^{2}+\left\Vert \psi\right\Vert ^{2}\right)_{0}^{2\pi}+2\left\Vert \phi\right\Vert ^{2}+12\left\Vert \psi\right\Vert ^{2}=\int_{0}^{2\pi}\gamma\phi\psi dt\label{eq:DampCond}
\end{equation}
Here we assumed that the driver $\beta$ is $2\pi$-periodic, which
suggests to introduce the following norm in the above equation
\[
\left\Vert \cdot\right\Vert ^{2}=\int_{0}^{2\pi}\left(\cdot\right)^{2}dt
\]
The sufficient condition for the asymptotic decay of $\phi$ and $\psi$
follows from eq.(\ref{eq:DampCond}):
\begin{equation}
\left\Vert \phi\right\Vert ^{2}+6\left\Vert \psi\right\Vert ^{2}>\frac{1}{2}\int_{0}^{2\pi}\gamma\phi\psi dt\label{eq:DamCond1}
\end{equation}
Applying then the Schwartz's and triangle inequalities to its r.h.s.,
we find 
\[
\int_{0}^{2\pi}\gamma\phi\psi dt\le\left\Vert \gamma\right\Vert \left\Vert \phi\psi\right\Vert \le\frac{1}{2\sqrt{6}}\left\Vert \gamma\right\Vert \left(\left\Vert \phi\right\Vert ^{2}+6\left\Vert \psi\right\Vert ^{2}\right)
\]
\[
\le2\left(\left\Vert \phi\right\Vert ^{2}+6\left\Vert \psi\right\Vert ^{2}\right)
\]

Hence, the stability condition is met if the driver is limited by:
\[
\left\Vert \gamma\right\Vert \le4\sqrt{6}
\]
Of course, the violation of this condition does not mean that the
instability sets on, but, if it does, the threshold will be very high
compared to the driver's amplitudes shown to be sufficient for a strong
excitation of even moments, as demonstrated in Secs.

\section{Random-Phase Approximation\protect\label{sec:Random-Phase-Approximation}}

In deriving the $\psi_{0}\left(t\right)$ growth shown in eq.(\ref{eq:psi0prime}),
it is convenient to use the variable $g=\beta\psi^{\prime}$ along
with $\psi$, so that eq.(\ref{eq:psi-sec-ord-eq}) takes the form

\begin{equation}
g^{\prime}+6g+4\beta^{\prime}\psi=0\label{eq:for-g-inApp}
\end{equation}
Now we decompose the involved variables into slowly varying and oscillating
parts:
\[
\psi=\psi_{0}+\tilde{\psi};\qquad g=g_{0}+\tilde{g};\qquad\beta=1+\tilde{\beta},
\]
cf. eqs(\ref{eq:betaSer}) and (\ref{eq:psiSer}). Retaining only
quadratic in the driver amplitude, $\tilde{\beta}$, terms, from eq.(\ref{eq:psi-sec-ord-eq})
we then have:
\[
g_{0}=\psi_{0}^{\prime}+\overline{\tilde{\beta}\tilde{\psi^{\prime}}}=\psi_{0}^{\prime}-\overline{\tilde{\beta^{\prime}\tilde{\psi}}};\qquad\tilde{g}\approx\psi_{0}^{\prime}\tilde{\beta}+\tilde{\psi}^{\prime}
\]
Using the above relation along with averaged eq.(\ref{eq:for-g-inApp}),
we obtain: 
\[
g_{0}^{\prime}+6g_{0}+4\overline{\tilde{\beta^{\prime}\tilde{\psi}}}=0;\qquad g_{0}^{\prime}+2g_{0}+4\psi_{0}^{\prime}=0
\]
so that for $\psi_{0}^{\prime}$ we can write:

\[
\qquad\psi_{0}^{\prime}\approx-\frac{1}{2}g_{0};
\]
Now ntroduce an auxilary function $\tilde{\lambda}$ as follows:
\[
\left(\tilde{\lambda}e^{6t}\right)^{\prime}=\tilde{\beta}^{\prime}e^{6t};\qquad\tilde{\lambda}=\sum\frac{i\omega_{k}\beta_{k}}{6+i\omega_{k}}e^{i\omega_{k}t}
\]
Then, from the oscillating part of eq.(\ref{eq:for-g-inApp}) we have:
\[
\left(\tilde{g}e^{6t}\right)^{\prime}+4\psi_{0}\left(\tilde{\lambda}e^{6t}\right)^{\prime}\approx0;\qquad\tilde{g}\approx-4\psi_{0}\tilde{\lambda}
\]
From $g=\beta\psi^{\prime}$ we get 
\[
\psi_{0}^{\prime}\approx g_{0}-\overline{\tilde{\beta}\tilde{g}}\approx g_{0}+4\psi_{0}\overline{\tilde{\beta}\tilde{\lambda}}=g_{0}+8\psi_{0}\sum\frac{\left|\beta_{k}\right|^{2}\omega_{k}^{2}}{36+\omega_{k}^{2}}
\]

\section{Floquet Analysis\protect\label{sec:Floquet-Calculations}}

Floquet theory is desined to study the solutions of linear systems
of differential equation with periodic coeffiecients. In the standard
textbooks, e.g., \citep{Coddington1955,Magnus2013}, it is explained
in more general terms than it is needed for the present derivation,
so it is easier to present it here in a simpler way. 

We fist rewrite eq.(\ref{eq:psi-sec-ord-eq}) in the following form
\begin{equation}
\left(\beta e^{6t}\psi^{\prime}\right)^{\prime}+\beta^{\prime}e^{6t}\psi=0,\label{eq:psi-sec-ord-eq-1}
\end{equation}
from which we obtain the following simple relation for the Wronskian,
$W\left(t\right)$:
\[
\left(\beta e^{6t}W\right)^{\prime}=0
\]
where the Wronskian is defined here as 
\[
W=\psi_{1}^{\prime}\psi_{2}-\psi_{2}^{\prime}\psi_{1}
\]
with $\psi_{1,2}$ being two linearly independent solutions of eq.(\ref{eq:psi-sec-ord-eq-1})
in one period $0<t<\tau$, satisfying the following initial conditions
at $t=0$:
\begin{equation}
\psi_{1}=\psi_{2}^{\prime}=1,\,\,\psi_{1}^{\prime}=\psi_{2}=0\label{eq:inCondPsi}
\end{equation}
so that 
\[
W\left(t\right)=\frac{\beta\left(0\right)}{\beta\left(t\right)}e^{-6t}
\]
If $\psi\left(t\right)$ is a solution of eqs.(\ref{eq:psi-sec-ord-eq})
and (\ref{eq:psi-sec-ord-eq-1}), then $\psi\left(t+\tau\right)$
must also be, because $\beta\left(t+\tau\right)=\beta\left(t\right)$.
Then we can find such a pair of constants $C_{1}$ and $C_{2}$ that
$\psi$ is a linear combination of $\psi_{1}$ and $\psi_{2}$, $\psi\left(t\right)=C_{1}\psi_{1}\left(t\right)+C_{2}\psi_{2}\left(t\right)$,
sutisfying the condition $\psi\left(t+\tau\right)=\rho\psi\left(t\right),$where
$\rho=const.$ For this to be true, the following conditions must,
in particular, be met for $t=0$. Then, according to eqs.(\ref{eq:inCondPsi})
we have

\begin{eqnarray}
C_{1}\psi_{1}\left(\tau\right)+C_{2}\psi_{2}\left(\tau\right) & = & \rho C_{1}\nonumber \\
C_{1}\psi_{1}^{\prime}\left(\tau\right)+C_{2}\psi_{2}^{\prime}\left(\tau\right) & = & \rho C_{2}\label{eq:FloqSyst}
\end{eqnarray}

This means that $\rho$ must belong to the spectrum of the Floquet's
fundamental solution matrix
\[
\mathcal{F}\left(t\right)=\left(\begin{array}{cc}
\psi_{1} & \psi_{2}\\
\psi_{1}^{\prime} & \psi_{2}^{\prime}
\end{array}\right)
\]
Its determinant $\left\Vert \mathcal{F}\right\Vert =-W\left(t\right)$
found above, while its spectrum $\rho_{1,2}$ (Floquet's multipliers)
comes from the solvability condition for eq.$\left(\ref{eq:FloqSyst}\right)$:
\[
\left\Vert \mathcal{F}-\rho\right\Vert =0
\]
From this equation the result shown in eq.(\ref{eq:FloquetMult})
follows. 

Finally we note that, more generally, we can write $\mathcal{F}\left(t+\tau\right)=\mathcal{F}\left(t\right)\cdot C\left(\tau\right)$
instead of eq.(\ref{eq:FloqSyst}), where $C$ is the so-called monodromy
matrix. Since $\mathcal{F}\left(0\right)=1$ (unit matrix) according
to eq.(\ref{eq:inCondPsi}), $C\left(\tau\right)=\mathcal{F}\left(\tau\right)$.
We have overridden this step, as we are only interested in the eigenvalues
$\rho$ of the matrix $C$. 

\bibliography{}

\end{document}